# Modulation of crystal and electronic structures in topological insulators by rare-earth doping

Zengji Yue*, Weiyao Zhao, David Cortie, Zhi Li, Guangsai Yang and Xiaolin Wang*

1. Institute for Superconducting & Electronic Materials, Australian Institute of Innovative Materials, University of Wollongong, Wollongong, NSW 2500, Australia

2. ARC Centre for Future Low-Energy Electronics Technologies (FLEET), University of Wollongong, Wollongong, NSW 2500, Australia

Email: zengji@uow.edu.au; xiaolin@uow.edu.au;

**Abstract**

We study magnetotransport in a rare-earth-doped topological insulator, $Sm_{0.1}Sb_{1.9}Te_3$ single crystals, under magnetic fields up to 14 T. It is found that that the crystals exhibit Shubnikov–de Haas (SdH) oscillations in their magneto-transport behaviour at low temperatures and high magnetic fields. The SdH oscillations result from the mixed contributions of bulk and surface states. We also investigate the SdH oscillations in different orientations of the magnetic field, which reveals a three-dimensional Fermi surface topology. By fitting the oscillatory resistance with the Lifshitz-Kosevich theory, we draw a Landau-level fan diagram that displays the expected nontrivial phase. In addition, the density functional theory calculations shows that Sm doping changes the crystal structure and electronic structure compared with pure $Sb_2Te_3$. This work demonstrates that rare earth doping is an effective way to manipulate the Fermi surface of topological insulators. Our results hold potential for the realization of exotic topological effects in magnetic topological insulators.

**Introduction**

Topological insulators (TIs) are one type of topological quantum matter, which have an insulating bulk state and a topologically protected metallic surface state with spin and momentum helical locking and a Dirac-like band structure.[1-2] A series of well-known thermoelectric semiconductors, including $Bi_2Se_3$, $Bi_2Te_3$, and $Sb_2Te_3$, have been identified as topological insulators materials.[3] Unique and fascinating electronic properties, such as the quantum spin Hall effect (QSHE), the quantum anomalous Hall effect (QAHE), topological magnetoelectric effects, and giant magnetoresistance (MR), are expected from topological insulator materials.[4-7] The backscattering by nonmagnetic impurities is also prohibited for the surface states due to their protection by time-reversal symmetry (TRS). Topological insulator materials also exhibit a number of excellent optical properties, including ultrahigh bulk refractive index values, near-infrared frequency transparency, unusual electromagnetic scattering, and ultra-broadband surface plasmon resonances.[8-10] These excellent electronic and optical properties enable topological insulator materials to be suitable for designing various advanced electronic and optoelectronic devices.[11-13] Magnetic doping can open up an energy gap at the Dirac point in topological insulators due to the breaking of time reversal symmetry by magnetic impurities.[14] A variety of exotic topological effects, including the quantum anomalous Hall effect (QAHE), topological magnetoelectric effects, image magnetic monopoles, and Majorana fermions, can be generated through magnetic doping. For example, the QAHE has been predicted and observed in transition-metal-doped magnetic topological insulators [5, 15]. The spontaneous magnetic moments and spin-orbit coupling in magnetic topological insulators can also lead to the QAHE without an external magnetic field. The QAHE demonstrates great potential for practical applications in next generation electronics with low power consumption. At present, however, the QAHE in magnetic topological insulators is limited to extremely low temperatures, down to 1 K. The realization of a high temperature QAHE is extremely desirable for both fundamental studies and practical applications.

Due to their large magnetic moment, rare-earth elements have been widely used in magnets and electronic devices. By breaking the time reversal symmetry, a magnetic moment can lead to the formation of a band gap in a topological insulator, where the size of the gap scales in proportion to the magnetic moment. As rare earths have some of the largest magnetic moments of any elements in the periodic table, and the 4$f$ electrons are heavily shielded, allowing for orbital moments, these could potentially be ideal for doping into TIs. Unlike defect induced magnetism, rare-earth (RE) element doping can also potentially generate intrinsic ferromagnetism.[16-17] Therefore, it is of interest to expand the search to identify promising candidate ferromagnetic topological insulators with wide band gaps, which are essential for the realization of a high temperature QAHE. Another promising avenue is to explore the proximity effects between rare-earth and transition-metal doped regions which are known in some cases to lead to enhanced temperatures.

Recent X-ray spectroscopy work has shown that the rare-earth dopants in topological insulators almost always take on the trivalent charge state.[18] With their ionic radii between 0.9 and 1.2 Å, and their trivalent states, the RE ions are a good match with Sb/Bi sites in the chalcogenide family. In thin films, high doping concentrations up to 30 % have been reported.[19] In single crystals, much lower peak doping concentrations have been reported, which are generally limited to < 2%. There may be exceptions to this limitation, however, because one group has reported crystalline $RE_xSb_{2-x}Te_3$ doped with as much as $x \sim 50\%$ for Ho, Sm, and Tb. Despite this promising set of materials, so far, detailed transport studies have not been performed to elucidate the effects of the rare earth on the strongly correlated electronic structure. The Sm ion is a particularly interesting case because both the 2+ and the 3+ ions are known to be stable. The 2+ version has a 4f6 configuration, and formally has $J = 0$ in its ground state, since the spin and orbital moments cancel according to Hund's rules. The first excited crystal-field split state of $Sm^{2+}$ is magnetic, however, and at low enough energy to be thermally occupied at hundreds of kelvins. The 3+ version of Sm has a 4f configuration that is intrinsically magnetic in its ground state, with $S = 5/2$ and $L = 3$. According to Russel-Saunders coupling, $J = L - S$, and the orbital contribution dominates the net moment for $Sm^{3+}$, which is typically < 1 Bohr magneton. Nevertheless, in practice, the magnetic moment of $Sm^{3+}$ is often anomalous and is in less than perfect agreement with the standard model of Hund's rules compared to the other RE ions. Mixed valence states of $Sm^{2+}/Sm^{3+}$ are also relatively common in practice, for example in $SmB_6$. The role of the 4f wavefunction in modifying the topologically protected states is currently the subject of much discussion on Sm-based systems. Consequently, the electronic structure of the Sm dopant in $Sb_2Te_3$ is of great interest.

Magnetotransport measurements represent an important approach to explore the electronic characteristics of quantum materials.[7, 20-24] In particular, the Shubnikov–de Haas (SdH) oscillations can probe the electronic structure, reveal information on the Fermi surface topology, and access the Berry phase.[25-26] SdH oscillations have also been used to investigate the dynamics of massless electrons in surface states.[27] The surface mobility and Fermi velocity of Dirac electrons can be obtained from the oscillation analysis. Here, we carried out magneto-transport measurements on Sm doped $Sb_2Te_3$ single crystals in different tilted magnetic fields. We observed SdH quantum oscillations that present the signature of a three-dimensional (3D) Fermi surface in $Sm_{0.1}Sb_{1.9}Te_3$. The Landau-level (LL) fan diagram reveals the nontrivial π

Berry phase. The rare earth doping can modify the crystal structure and electronic structure, and is also promising for achieving novel topological magneto-electronic effects.

**Methods:**

High-quality $Sm_{0.1}Sb_{1.9}Te_3$ single crystals were grown using the melt-cooling method, starting with high-purity Sm, Sb, and Te elements. The elements were sealed in a quartz tube, heated to 700 ºC, and kept for 6 hours. Then, the temperature was reduced to 650 ºC and slowly cooled down to 550 ºC at the rate of 2 ºC/h. Finally the temperature was allowed to decrease down to room temperature naturally. The quality of the bulk crystals was investigated using X-ray diffraction (XRD), and the growth orientation was found to be along the [001] direction. Density functional theory (DFT) calculations were carried out using the Vienna Ab-initio Simulation Package (VASP) version, 5.44 based on the plane-augmented wave (PAW) methodology.[28-31] All calculations were performed using the Perdew-Burke-Ernzerhof generalized gradient approximation (PBE GGA) functional.[32] Spin polarization and orbital polarization were included self-consistently, together with the spin orbit interaction[33] Both pure $Sb_2Te_3$ and $Sm_xSb_{2-x}Te_3$ were simulated to the same levels of precision. The energy cut-off and electronic convergence values were 300 eV and $1.0E10^{-5}$ eV, respectively. Forces were converged to within 0.02 eV/Å. Dispersion corrections were included via the Grimme D3 method to account for van der Waals interactions.[34] A k-point mesh equivalent to a 16×16×2 mesh in the hexagonal unit cell was adopted. The k-point integration was conducted using the tetrahedron method with Blöchl corrections. For the $Sm_xSb_{2-x}Te_3$, a 2×2×1 hexagonal supercell was constructed containing a single Sm atom and 23 Sb atoms, corresponding to $x = 0.04$. Smaller cells with compositions closer to experimental values were also investigated, but these unavoidably introduced spurious self-interactions between the 4f dopants due to direct Sm-Sm in-plane nearest-neighbor interactions across the periodic boundary conditions, and thus, the larger cells are taken as a better approximation to the experimental system. The correct *ab-initio* treatment of 4f electrons is one of the grand challenges in computational quantum chemistry because the behavior of 4f electrons in solids is complicated by the interplay of hybridization, the sizable electronic exchange, crystal fields, and spin-orbit coupling.[35] While DFT is an important tool, it suffers from a well-known tendency to delocalize 4f electrons, and it frequently predicts metallic conductivity in 4f materials that are experimentally insulating. Thus, to model the Sm 4f electrons, we explored two complementary DFT-based approaches. In the first approach, the 4f electrons are treated as non-valence states and frozen in the core by using a $Sm^{3+}$ plane augmented wave potential (PAW). This treatment is known to give results that are comparable to both the Hubbard I mode and the "standard model" of 4f electron systems, thus giving reliable results for crystal structures and volumes in insulators.[36] The second approach used was to treat the 4f electrons as valence electrons, but apply the GGA + Hubbard correction (GGA+U) formalism, which is the standard approach for strongly-correlated electron systems. The calculations in this second approach apply strong on-site Coulomb interactions ($U$ = 6 eV) and on-site exchange ($J$ = 1 eV) to the 4f orbitals for Sm in $Sb_2Te_3$. These values are consistent with the high values known for strongly-correlated Sm systems ($U$ = 5-7 eV).[35] Investigations with different values of $U_{eff} = U - J$ were performed for a range of values and found to yield similar results, provided $\underline{U}_{eff}$ > 4 eV. Similar approaches have been used elsewhere to model rare-earth metals and dopants.[35]

**Results:**

Figure 1a shows the XRD pattern of the $Sm_{0.1}Sb_{1.9}Te_3$ single crystals, which demonstrates the high quality of the crystals. Several small pieces of crystal were cleaved from the bulk crystals for the transport measurements, as shown in the inset of Fig. 1b. In the transport measurements, we used the standard four-probe configuration with the current applied in-plane. The measurements were conducted in a Physical Properties Measurement System (PPMS, Quantum Design). The angular dependence of the magnetoresistance (MR) was measured using a standard horizontal rotation option in the PPMS. The tilt angle $\theta$ between $B$ and the $a$-axis can be varied from 0 to 90°. Figure 1b shows the temperature dependence of the resistance in $Sm_{0.1}Sb_{1.9}Te_3$ single crystals, which presents a metallic character. Ideally, topological insulators should have insulating bulk states and metallic surface states. The measured conductance should come entirely from the surface states. In real topological insulator materials, however, the surface conductance is always masked by the conductance from non-insulating bulk states, which are caused by bulk defects.[37] The Fermi energy generally falls either in the conduction band or the valence band rather than in the expected bulk band gap.[38-39] In $Sm_{0.1}Sb_{1.9}Te_3$, the Fermi surface is located in the valence band and the majority carriers are holes, which leads to its high bulk conductance.

Magnetoresistance (MR = $(R_H-R_0)/R_0 \times 100\%$) is the change of electrical resistance of a material in an externally-applied magnetic field ($R_H$) compared to the resistance at zero field

($R_0$). It has wide applications in magnetic data storage, magnetic sensors, and magnetoelectronic devices.[40-41] Figure 1c shows the MR in $Sm_{0.1}Sb_{1.9}Te_3$ crystals at different temperatures. The MR shows quantum oscillations at low temperatures and high magnetic fields. The MR values reach as high as 160% at 3.5 K and 14 T. We previously studied the MR in undoped $Sb_2Te_3$ bulk crystals.[42] Giant MR of up to 230% was observed in Bi doped $Sb_2Te_3$ crystals, which exhibit a quadratic character in low fields and a linear characteristic at high fields. The giant MR also shows strong anisotropy, with the anisotropy ratio as high as 210% in angle-dependent measurements. This means that the Sm doping changes the electron density and mobility in this compound. Figure 1d shows the Hall resistance in $Sm_{0.1}Sb_{1.9}Te_3$ crystals at different temperatures. The measurements of both longitudinal resistivity and Hall resistivity are important, not only for calculating the carrier density and Hall mobility, but also for assigning the LL index and determining the Berry phase. Based on the Hall effect and results, we calculated the carrier density and mobility. The Hall coefficient is $R_H = 1.8 \times 10^{-2}$ cm$^3$/C. The carrier density $n = 3.2\times10^{18}$. The carrier mobility is $\mu = 2.52 \times 10^3$ cm$^2$V$^{-1}$s$^{-1}$. Table 1 compared the parameters between pure $Sb_2Te_3$ and $Sm_{0.1}Sb_{1.9}Te_3$ crystals.[26]

Quantum oscillations are widely used to understand the band structure of topological insulators, Rashba materials, and Dirac semimetals.[25, 43] Measurements that are dependent on the magnetic field orientation make it possible to identify the topological surface states. We observed quantum oscillations in $Sm_{0.1}Sb_{1.9}Te_3$ crystals and studied the dependence of quantum oscillations on the orientation of the magnetic field. Figure 2a shows the magnetic field dependence of the resistance at different angles at 3.5 K. At low temperature, the resistance curves display evident quantum oscillations. The quantum oscillation patterns result from the SdH effect. To analyse the SdH oscillations, we subtract the resistance background. Figure 2b shows the SdH oscillation patterns obtained by subtracting the smooth backgrounds at different angles. Some different shapes of harmonics appear periodically. Figure 2c shows the amplitudes of the fast Fourier transform (FFT) of the oscillation patterns. The peaks are located at 5, 20, 25, and 45 T. Some peaks disappear with the degree changes from 90° to 0°, which means the surface contribution disappears at 0°. As some peaks survive to 0°, the SdH oscillations result from both bulk states and surface states.

To study the topological and Berry phases, we need fit the oscillatory resistance with the Lifshitz-Kosevich theory. We selected the data at 3.5 K and three different degrees, 0°, 45°, and

90°. Figure 3a shows the SdH oscillation patterns at the selected angles at 3.5 K. Figure 3b shows the Landau level (LL) index diagram of $Sm_{0.1}Sb_{1.9}Te_3$ single crystals and the values of the intercepts of the fitting lines. The extrapolated Landau-level index $\nu$ at the extreme field limit, e.g., $1/B \rightarrow 0$, is related to Berry's phase, which indicates a phase shift regarding the conventional Landau quantization in materials. The intercept $\nu = 0$ corresponds to a normal metal, while $\nu = 0.5$ comes from the massless relativistic fermions in a magnetic field. SdH oscillations originate from successive emptying of Landau levels (LL) with increasing magnetic field. The LL index $n$ is related to the cross-sectional area $A_F$ of the Fermi surface by $2\pi(n + \gamma) = A_F\hbar/eB$ where $\gamma = 0$ or $1/2$ for topologically trivial electrons and Dirac fermions, respectively, $e$ is the electron charge, $\hbar$ is the reduced Planck's constant ($\hbar = h/2\pi$), and $B$ is the magnetic field. As shown in Fig. 3b, the intercept for the $Sm_{0.1}Sb_{1.9}Te_3$ single crystal is 0.3, indicating that the quantum oscillations might be contributed by a mixed contribution of surface states and bulk states in $Sm_{0.1}Sb_{1.9}Te_3$ due to strong spin orbital coupling in this system.

**Discussion**

SdH oscillations have been observed in a number of Dirac materials, including graphene, topological insulators, and 3D materials.[44-46] When a crystal is located inside a magnetic field, electrons engage in cyclotron motion and are quantized into a series of Landau levels. With the magnetic field increasing, the Landau levels move to higher energy, and the electrons become free to flow as current when each energy level passes through the Fermi energy. This leads to periodical oscillations in the conductivity, which are called SdH oscillations. SdH oscillations can be used to map the Fermi surface of electrons in a crystal, by determining the periods of oscillation for various applied field directions. They can also be used to determine the effective mass of charge carriers, allowing us to distinguish between majority and minority carrier populations. We observed the SdH oscillations in rare-earth doped topological insulators. These observations demonstrate an important signature for the existence of the 3D Fermi surface. The oscillation frequency $F$ is related to the cross-section of the Fermi surface $A$ by the Onsager relation: $F = (\hbar c/2\pi e)\cdot A$; here, $\hbar = h/2\pi$, where $h$ is Planck's constant, and $e$ is the elementary charge.[47] Therefore, the cross-section $A$ of $Sm_{0.1}Sb_{1.9}Te_3$ is 0.32 $nm^{-2}$ and 0.79 $nm^{-2}$. The relative Fermi wave vector $K_F$ can be calculated as $K_F = (A/\pi)^{1/2}$ giving 0.32 $nm^{-1}$ and 0.5 $nm^{-1}$. The density $n$ is related to the Fermi wave vector $K_F$ by $n = K^2_F/2\pi$. According to Lifshitz-Kosevich (LK) theory, $M_{os} \propto R_T R_D \sin(\frac{2\pi F}{B} + \beta)$, where $M_{os}$ is the magnitude of the oscillation, $R_T$ is the temperature damping factor, $R_D$ is the Dingle damping factor, and $\beta$ is the Berry phase, respectively. The effective mass $m^*$ can be extracted from the temperature dependence of the SdH oscillation amplitudes by $R_T = \frac{\alpha T m*}{B \cdot \sinh(\alpha T m*/B)}$, in which $\alpha = \frac{2\pi^2 k_B m_e}{eh} \sim$ 14.96 T/K, where $k_B$ is the Boltzmann constant and $m_e$ is the electron rest mass. We fitted the temperature dependence of the SdH oscillation amplitude of $Sm_{0.1}Sb_{1.9}Te_3$ to obtain the effective mass of 0.28 $m_e$.

To analyse the possible origin of the Fermi surface features obtained from the SDH oscillations, density functional theory (DFT) calculations were conducted. These reveal a complex interplay between the electronic and the crystal structure. Pure $Sb_2Te_3$ possesses a rhombohedral crystal structure that can be visualized as a layered structure with a hexagonal lattice cell. The layered structure consists of two atomic sheets of Sb and three atomic sheets of Te, held together by weak van der Waals forces.[48] Given their similar valence state and ionic radii, Sm is expected to substitute on Sb sites. Figure 4a shows the crystal structure obtained from DFT for a Sm dopant in $Sb_{2-x}Te_3$ viewed along the [001] direction after ionic relaxation of the atomic positions and cell parameters. A cross-sectional view of the same crystal structure along the [100] direction is shown in Figure 4b. The spin difference is superimposed on the crystal lattice using the colored section, thus demonstrating that the moment is localized around the Sm dopant. To clarify the local crystallographic distortion of the Sm dopant, Figure 4c shows an enlarged region around a Sb atom in its octahedral coordination compared with the Sm environment (Figure 4d). In pure $Sb_2Te_3$, the DFT calculation indicates that the Sb-Te octahedra contain a mixture of long (3.17 Å) and short bonds (3.02 Å). Examining the published experimental crystal structure data indicates that the former bond lengths are within 1% of the experimental data, and the shorter bond lengths reflect the mutual Te-Te repulsion across the van der Waals gap.[49] Samarium doping modifies this flexible octahedral environment by enforcing two sets of elongated, equidistant bonds (3.2 Å), which has the effect of pushing the Sm slightly out of the plane (Figure 4b). The basic qualitative result is independent of whether the frozen core or GGA+U method is used. Thus a significant local crystal structure distortion is introduced by the Sm dopant, indicating that the Sm-Te octahedra are more rigid and less distorted than the surrounding Sb-Te octahedra.

The *ab initio* calculations reveal that the electronic structure is sensitive to the presence of samarium dopant. There are two independent mechanisms by which a rare-earth dopant may modify the electronic structure of the $Sb_2Te_3$. Firstly, the samarium dopant introduces crystallographic distortions that may affect the delicate semimetallic Sb-Te bands. Secondly, the 4*f* electrons from the samarium dopant may potentially form entirely new bands, which can potentially hybridize with *s/p* Sb-Te bands in specific regions across the Fermi surface. The first effect is generally expected to be important in $Sb_2Te_3$ because the material is known to be very sensitive to lattice distortions. To illustrate this point, Figure 5a shows the band structure of pure $Sb_2Te_3$ compared with the experimental lattice constants (dashed lines) and the cell parameters derived by DFT (lines). Although the predicted and experimental values are within 1% of each other, even a small degree of strain in the $Sb_2Te_3$ has strong effects on the band structure of the semi metal. In particular, cell distortions alter the valence band offset between the light hole and heavy hole bands with respect to the $\Gamma$ point band. Consequently, the larger crystallographic distortions from the Sm ion induce noticeable effects on the band structure of $Sb_2Te_3$. To isolate the effect of crystal structure distortion from those of 4*f* valence contributions, we compare the frozen-core calculations with the calculations where the 4*f* orbitals are treated as valence orbitals in Figure 5b and c). Clearly the distortion introduced by Sm alone is sufficient to modify the valence band offsets near the $\Gamma$ point, as shown in Figure 5b. In particular, the offset between the $\Gamma$ point bands and the M point hole bands is reduced owing to the Sm dopant. Experimentally, this may explain the additional frequency observed in the experimental Shubnikov-de Haas oscillations, although the details depend on the precise position of the Fermi level in the valence band, which, in turn, depends on the other intrinsic defects in the $Sb_2Te_3$. A very different electronic structure emerges in the calculations where Sm are treated as valence (Figure 5c) because the spin-polarized 4*f* states in the calculation introduce many additional complex flat bands (red color) near the Fermi level, in combination with a modified $\Gamma$-M offset. Furthermore, there is a significant shift of the Fermi level for the GGA+U + SO calculations, where SO is spin-orbit, indicating the tendency for Sm to act as an acceptor dopant in the calculation, shifting the Fermi level deeply into the valence band. This is confirmed by examining the 4*f* charge occupation, which is close to 4*f* in the calculation, indicating a preference for the $Sm^{2+}$ valence state. Based on the experimental evidence, however, given the small change in the Hall resistivity, it appears that the second scenario is unrealistic, and the frozen-core scenario is in better agreement with experiment. Furthermore, experimental magnetic measurements demonstrated a paramagnetic $Sm_{0.1}Sb_{1.9}Te_3$ property down to 2 K, which is inconsistent with the $Sm^{2+}$ non-magnetic valence state and in agreement with the frozen core $Sm^{3+}$ scenario.

The goal of ferromagnetic doping in topological insulators is to induce a Dirac-mass gap in the topological surface state band structure at the Dirac point. The rare earth elements have high magnetic moments and are often used for producing rare-earth magnets. Ho, Gd, and Dy have been used as dopants for $Bi_2Te_3$ thin films, but no long-range ferromagnetic order has been detected.[50-52] A gapped surface state band has been observed in Dy-doped $Bi_2Te_3$.[53] Short-range ferromagnetic order exists in this system, which results from inhomogeneous doping and the formation of clusters and defects. Our $Sm_{0.1}Sb_{1.9}Te_3$ demonstrates the paramagnetic property down to 2 K based on our magnetization measurements. Higher doping may be needed for obtaining long-range ferromagnetic order. Further theoretical studies are expected for gaining a better understanding of rare earth doping in topological insulators.

**In summary**

In summary, comprehensive magneto-transport measurements have been conducted on rare earth Sm doped $Sb_2Te_3$ single crystals. We observed large positive MR of up to 160% and high-field quantum oscillations, which demonstrate a 3D Fermi surface in Sm doped $Sb_2Te_3$. The fan diagram shows that the Fermi surface has a nontrivial π Berry phase. Angle-dependent SdH oscillations demonstrated that the measured conductance was derived from both bulk states and surface states. The rare earth Sm doping changed both the crystal structure and the electronic structure. Our work demonstrates that the tuning of electron transport by rare-earth magnetic doping paves the way to finding exotic topological effects.

Table 1

| TIs | $m$ | $\mu$ ($cm^2V^{-1}s^{-1}$) | $n$ ($cm^{-3}$) | $R_H$ ($cm^3/C$) | Berry phase |
|---|---|---|---|---|---|
| $Sb_2Te_3$ | 0.23 $m_e$ | $2.16 \times 10^3$ | $9.5 \times 10^{19}$ | $6.5 \times 10^{-2}$ | 0.5 |
| $Sm_{0.1}Sb_{1.9}Te_3$ | 0.28 $m_e$ | $2.52 \times 10^3$ | $3.2 \times 10^{18}$ | $1.8 \times 10^{-2}$ | 0.3 |

Table 1 Comparation of the parameters between pure $Sb_2Te_3$ and $Sm_{0.1}Sb_{1.9}Te_3$ crystals.

**Figures**

Figure 1

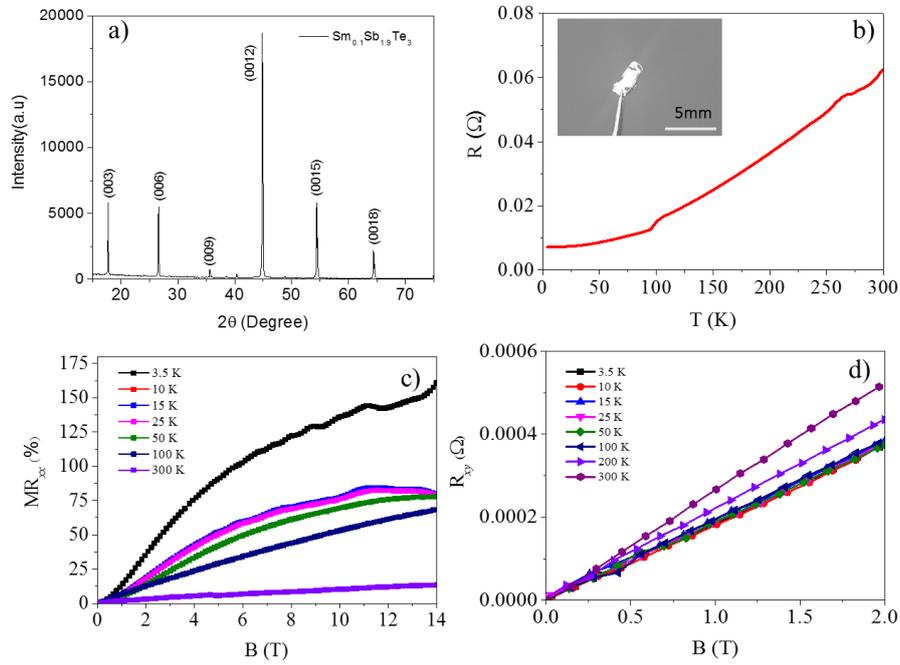

Figure 1. Characterization of the $Sm_{0.1}Sb_{1.9}Te_3$ single crystals. (a) X-ray diffraction pattern of $Sm_{0.1}Sb_{1.9}Te_3$ single crystals. (b) The temperature dependence of the resistance in $Sm_{0.1}Sb_{1.9}Te_3$ single crystals. The inset displays a photograph of a single crystal. (c) MR in $Sm_{0.1}Sb_{1.9}Te_3$ single crystals at different temperatures. (d) Hall resistances in $Sm_{0.1}Sb_{1.9}Te_3$ single crystals at different temperatures.

Figure 2

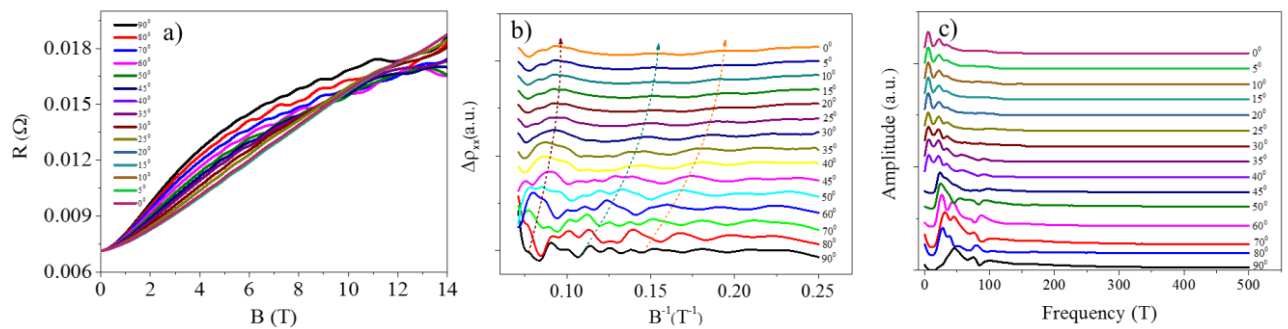

Figure 2. Magnetic field and angle dependent SdH oscillations in $Sm_{0.1}Sb_{1.9}Te_3$ single crystals. (a) The typical SdH oscillations without background subtraction. (b) SdH oscillations

as functions of 1/*B* measured from 0° to 90° at 3.5 K. (c) Fast Fourier transform spectra of SdH oscillation frequency from 0° to 90°.

Figure 3

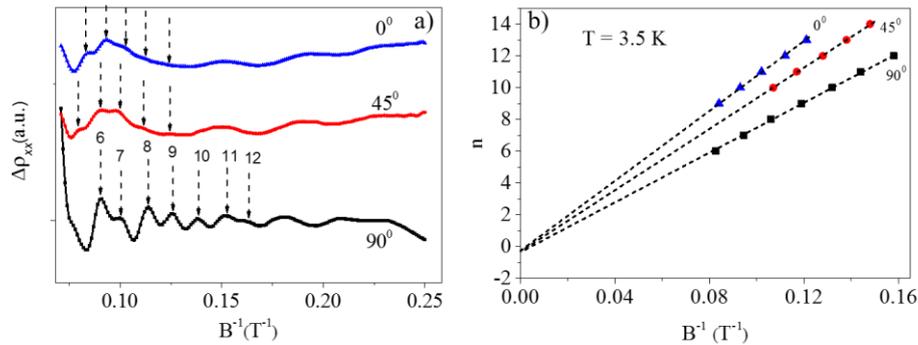

Figure 3. Analysis of the SdH oscillations in $Sm_{0.1}Sb_{1.9}Te_3$ single crystals. (a) Selected SdH oscillations at different angles labeled with the Landau number. (b) Landau level (LL) index diagram of $Sm_{0.1}Sb_{1.9}Te_3$ single crystals.

Figure 4

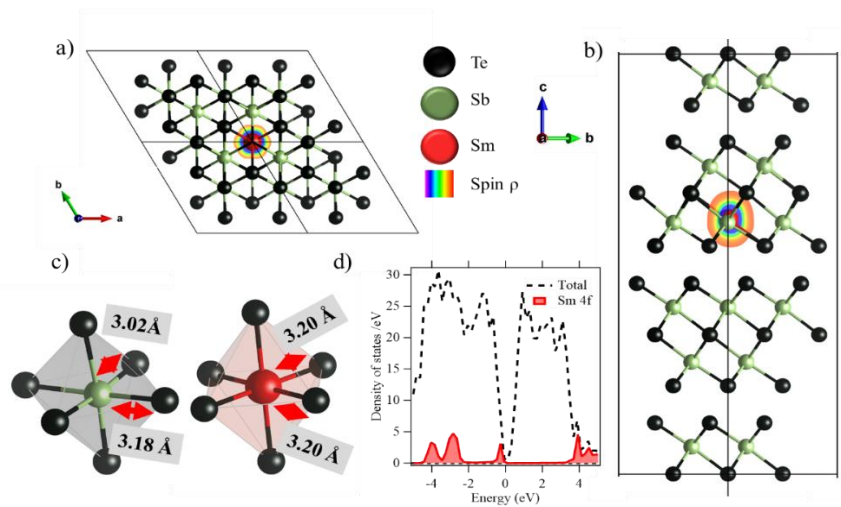

Figure 4. DFT calculations of the crystal structure and spin density of of $Sm_{0.04}Sb_{1.96}Te_3$ single crystals for supercell calculations after ionic relxaation (a) The Sm dopant (red) is positioned in a substitutional site, as viewed along the [001] axis. The dark atoms are Te, the light are Sb. The coloured gradient regions show the local spin density (b) The cell and spin density viewed along a (100) basal plane direction. (c) Both the Sb and

Sm are octahedrally coordinated, although the Sm introduces bond-length distortion into the cell d) The total electronic density of states and the Sm 4f contribution to the GGA+U+SO calculations.

Fig. 5

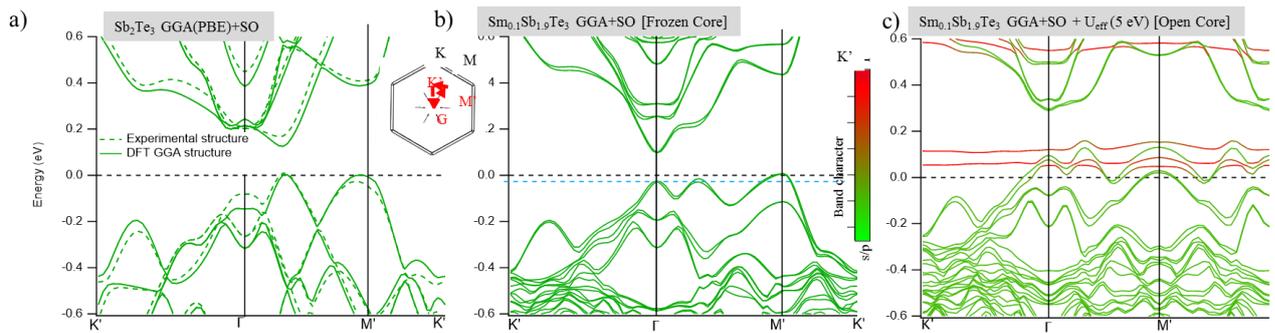

Figure 5. Band-structures calculated using DFT for a) pure $Sb_2Te_3$ using the experimental lattice constants (dashed lines) with the cell optimised using the GGA-PBE functional (solid lines). The inset shows a schematic illustration of the Brillouin zone. b) The band structure of $Sm_{0.04}Sb_{1.96}Te_3$ using the frozen core approximation for the $Sm^{3+}$ ion. c) $Sm_{0.04}Sb_{1.96}Te_3$ including 4f valence electrons according to the GGA+U+SO method. The color scale is the band-character plot with the color code reflecting the 4f versus *s/p* contribution for the calculated bands.